\documentclass[reprint,prl,twocolumn,nofootinbib,aps,floatfix]{revtex4-2}
\usepackage{amsmath}
\usepackage{amsfonts}
\usepackage{amssymb}
\usepackage{mathtools}
\usepackage{graphicx}
\usepackage[dvipsnames]{xcolor}
\usepackage[colorlinks=True,citecolor=myRed,linkcolor=myRed,urlcolor=gray]{hyperref}
\usepackage{hypcap}
\usepackage{braket}
\usepackage{yfonts}
\usepackage{bm}
\usepackage[caption=false]{subfig}

\usepackage{soul}
\usepackage{dsfont}
\usepackage{dblfloatfix}
\newcommand\eq[1]{\begin{align}#1\end{align}}

\newcommand\nh{N_\mathcal{H}}
\newcommand{\Ll}{\mathcal{L}}

\definecolor{myBlue}{RGB}{31,119,180}
\definecolor{myOrange}{RGB}{255,127,14}
\definecolor{myGreen}{RGB}{44,160,44}
\definecolor{myRed}{RGB}{214,39,40}
\definecolor{myPurple}{RGB}{148,103,189}

\makeatletter
\def\p@figure{\color{myBlue}}
\def\p@equation{\color{myRed}}
\makeatother

\begin{document}

\title{Spectral Multifractality and Emergent Energyscales Across the\\ Many-Body Localisation Transition}

\author{Sthitadhi Roy}
\email{sthitadhi.roy@icts.res.in}
\affiliation{International Centre for Theoretical Sciences, Tata Institute of Fundamental Research, Bengaluru 560089, India}

\begin{abstract}
We present a scaling theory of the many-body localisation transition in terms of emergent, characteristic energyscales. The analysis is based on the decomposition of the eigenstates in the basis of trivially localised states, resolved in the energies of the latter, which we refer to as the spectral decomposition of the eigenstates. The characteristic energyscales emerge when the multifractal properties, or lack thereof, of the spectral decomposition are studied at different scales. These characteristic scales correspond to the ones, above which the spectral decompositions exhibit their global behaviour, namely full ergodicity in the ergodic phase and multifractality in the many-body localised phase. On the other hand, at scales below the characteristic ones, the decomposition in the ergodic phase shows finer (multi)fractal structures whereas in the localised phase, the decomposition picks out well-separated, localised resonant peaks. The scaling of these characteristic energyscales across the many-body localisation transition admits a scaling theory consistent with a Kosterlitz-Thouless type scenario and bears striking resemblances to that of inverse participation ratios of eigenstates.
\end{abstract}

\maketitle

The nature of the many-body localisation transition (MBLT) has been a question of enduring interest in condensed matter and statistical physics for the better part of the last two decades~\cite{basko2006metal,gornyi2005interacting,oganesyan2007localisation,nandkishore2015many,abanin2019colloquium,sierant2024manybody}. Contrary to conventional, ground state quantum phase transitions~\cite{sachdev2011quantum}, the MBLT occurs at the level of individual eigenstates at arbitrary energy densities~\cite{oganesyan2007localisation,pal2010many}.
This poses a fundamental challenge as it is not {\it a priori} obvious that the traditional tools of equilibrium statistical mechanics continue to be effective in understanding the MBLT. This naturally has led to considerable debate regarding the nature of the MBLT with results ranging from those suggesting the MBLT to be a continuous transition~\cite{pal2010many,luitz2015many,potter2015universal,vosk2015theory,dumitrescu2017scaling,sierant2023stability} but with critical exponents often violating fundamental bounds~\cite{chayes1986finite,chandran2015finite} to Kosterlitz-Thouless-like scenarios~\cite{thiery2018many-body,goremykina2019analytically,dumitrescu2018kosterlitz,morningstar2019renormalization,morningstar2020manybody,mace2019multifractal,roy2021fockspace,sutradhar2022scaling}.

This is in large part due to the inherent difficulty with defining microscopic correlation/localisation lengthscales and energyscales that may exhibit universal scaling at the transition. 
Owing to the fundamentally interacting and many-body nature of the systems, there is no obvious correlation or localisation length such as those possessed by single particle wavefunctions in the context of Anderson transitions~\cite{evers2008anderson}. 
However, interpreting the many-body eigenstate wavefunctions as those of a fictitious single particle on the Fock/Hilbert space~\cite{logan1990quantum,altshuler1997quasiparticle} has allowed for identifications of volumescales and lengthscales on the Fock/Hilbert-space graph, analogous to those for Anderson localisation in high-dimensional graphs~\cite{garciamata2017scaling,garciamata2020two,garciamata2022critical}.
Scaling theories of the MBLT based on such quantities have provided compelling evidence in favour of the scenario that there exists a Fock-space correlation volume in the ergodic phase which diverges with an essential singularity as the transition is approached whereas there exists a Fock-space localisation lengthscale in the many-body localised (MBL) phase which approaches a finite value as the MBLT is approached from the MBL side~\cite{mace2019multifractal,roy2021fockspace}. 
These results do provide numerical credence to the Kosterlitz-Thouless-like scenario for the MBLT as predicted by phenomenological renormalisation group approaches~\cite{thiery2018many-body,goremykina2019analytically,dumitrescu2018kosterlitz,morningstar2019renormalization,morningstar2020manybody}.

{ However, the complete description of any quantum phase transition requires the universal properties of lengthscales {\it as well as energyscales} near the transition~\cite{sachdev2011quantum}, and the MBLT is no exception. Despite the fundamental importance of the question, an unambiguous identification of microscopical energyscales that may show universal scaling near the MBLT, and their scaling behaviour, has so far eluded us. Answering this question constitutes the central motivation of this work. }

% However, the MBLT being a dynamical quantum phase transition, it is of fundamental importance to identify energyscales that show universal properties at the transition; the relevant scaling behaviour along with the scaling behaviour of the aforementioned correlation volume- and lengthscales on the Fock/Hilbert-space graph can then be used to define analogues of dynamical exponents for the MBLT. 
% This constitutes the central motivation of this work. We address this precisely by identifying relevant energyscales on both sides of the MBLT and their universal properties near the transition.

In order to do so, we spectrally decompose the eigenstates in the basis of trivially localised states and analyse their finer structures in terms of the multifractal properties of the decomposition at different energyscales. 
In the ergodic phase, this yields a characteristic energyscale $\omega_\ast$, proportional to the mean level spacing, or equivalently, inversely proportional to the Hilbert-space dimension, $\omega_\ast \sim \nh^{-1}$. 
Probing the spectral decomposition below this energyscale shows an underlying multifractal nature whereas fully developed ergodicity is recovered upon probing the decomposition above the characteristic scale. 
On the other hand, in the MBL phase, there exists a different characteristic scale which scales anomalously with the Hilbert-space dimension, $\omega_\ast\sim \nh^{-\mu}$ with $0<\mu<1$. In this case, at energyscales smaller than this characteristic scale, the decomposition picks up the individual localised resonances whereas probing the decomposition at scales larger than the characteristic scale reveals the fully developed (multi)fractality of MBL eigenstates~\cite{deluca2013ergodicity,luitz2015many,mace2019multifractal}. Identification of these characteristic energyscales in the ergodic as well as in the MBL phase is the first of the two central results of this work.

The second and final central result of this work is the scaling of the aforementioned energyscales as the MBLT is approached from either side. In the ergodic phase, $\omega_\ast =\Lambda_\omega \nh^{-1}$ with $\Lambda_\omega$ diverging as an essential singularity as the MBLT is approached. On the other hand as the MBLT is approached from the MBL phase, $\mu\to \mu_c <1$, and $\mu$ approaches $\mu_c$ as a power-law. This is again indicative of the MBL critical point being a part of the MBL phase itself, consistent with earlier studies~\cite{thiery2018many-body,goremykina2019analytically,dumitrescu2018kosterlitz,morningstar2019renormalization,mace2019multifractal,morningstar2020manybody,roy2021fockspace,laflorencie2020chain,sutradhar2022scaling}. It is also interesting to note that the scaling of these energyscales across the MBLT is very similar to that of the inverse participation ratios (IPRs)~\cite{mace2019multifractal,roy2021fockspace}, and hence, one expects a scaling theory again similar to that of the IPRs. We find that our numerical results are consistent with this expectation, which concludes the main message of the work.

\begin{figure}\includegraphics[width=\linewidth]{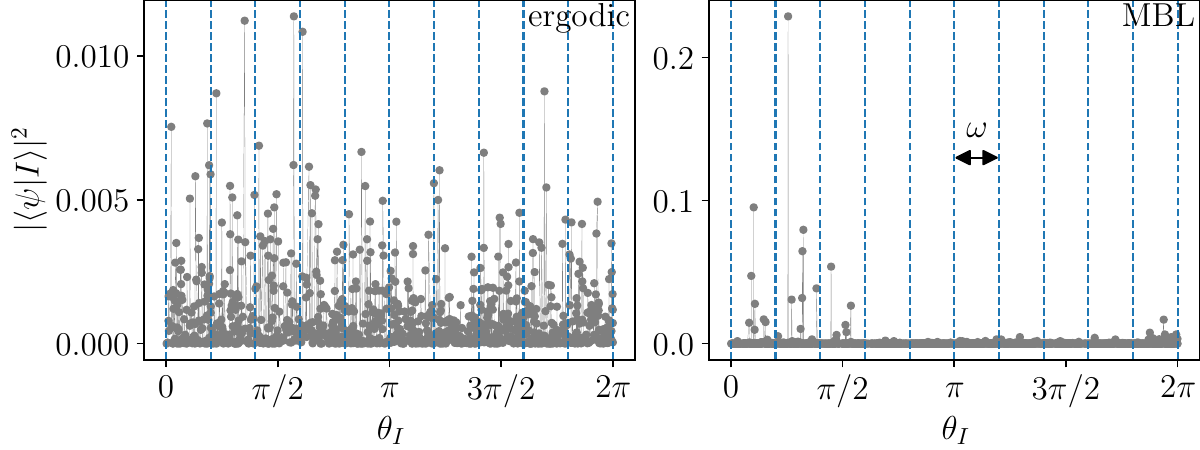}
\caption{Illustrative examples of the spectral decomposition of the ergodic (left) and MBL (right) eigenstates in the basis of the trivially localised states. The vertical dashed lines provide a graphical representation of the quasienergy segments (of width $\omega$) used to define the energy-dependent generalised IPR in Eq.~\ref{eq:Lq}. The data is for a Floquet spin-chain described in Eq.~\ref{eq:HX-HZ} with $L=10$.}
\label{fig:spectral-decomp}
\end{figure}

Before delving into the specifics, let us describe the approach towards extracting the energyscales in a general setting. The class of models we will be working with can be described in terms of a Hamiltonian $(H)$ or a Floquet unitary $(U_F)$ of the form
\eq{
H = H_\mathrm{loc} + \Gamma H_\mathrm{deloc};\quad U_F = e^{i H_\mathrm{loc}}e^{i \Gamma H_\mathrm{deloc}}\,,
\label{eq:H-UF-gen}
}
where $H_\mathrm{(de)loc}$ favours (de)localisation and the relative strength of two terms, $\Gamma$, drives the MBLT. 
It is more convenient to work with Floquet models as all  eigenstates are statistically equivalent and the density of quasienergies on the unit circle is uniform. Also, the finite-size effects in such models are arguably less severe~\cite{sierant2023stability,sierant2024manybody}. 

As a matter of convention, we denote the localised eigenstates of $e^{i H_\mathrm{loc}}$ as $\ket{I}$ with $e^{i H_\mathrm{loc}}\ket{I} = e^{i\theta_I}\ket{\theta_I}$, and the eigenstates of the full unitary as $\ket{\psi}$. The spectral decomposition of the eigenstate $\ket{\psi}$ is then defined as
\eq{
D(\theta) = \sum_{I=1}^{\nh} |\braket{\psi|I}|^2\delta(\theta\textbf{}-\theta_I)\,.
\label{eq:spectral-decomp}
} 
In Fig.~\ref{fig:spectral-decomp}, we show illustrative examples of the decomposition in both the ergodic and MBL phases. 
While the global features of the decomposition are as one expects,
% for ergodic and MBL eigenstates, 
the key point is, there is visible richer structure when the decomposition is viewed at finer energyscales. 
The decomposition at different energyscales is probed in the following way. 
We divide the quasienergies $\theta_I \in [0,2\pi)$ into segments of width $\omega$ and define a generalised $\omega$-dependent IPR as 
\eq{
\Ll_q(\omega) = \sum_{n=1}^{n_\mathrm{seg}}P_n^q(\omega)\,;\quad P_n(\omega)=\int_{(n-1)\omega}^{n\omega}d\theta~ D(\theta)\,,
\label{eq:Lq}
}
where $n_\mathrm{seg} = 2\pi/\omega$ is the number of segments. Physically, the quantity $P_n(\omega)$ contains the total weight of the eigenstate inside the $n^{\rm th}$-segment and $\Ll_q(\omega)$ can be interpreted as a generalised IPR of these weights over all the segments. 
The limiting behaviours of $\mathcal{L}_q(\omega)$ can be deduced easily. 
For $\omega=2\pi$, there is only one segment with the entire weight of the normalised eigenstate in that one segment, such that, $\mathcal{L}_q(2\pi) = 1$. 
For $\omega \ll 2\pi/\nh$, each segment has at most one $\theta_I$ such that $\mathcal{L}_q(\omega\to 0) = \sum_I |\braket{\psi|I}|^{2q}\sim \nh^{-\tau_q}$ which is nothing but the standard IPR, $\mathcal{I}_q$, of the eigenstate with $\tau_q$ the fractal exponent. 
{Two brief remarks are in order here. First, note that the conventional IPR is only a limiting behaviour of ${\cal L}_q(\omega)$ with the latter containing much finer grained information not accessbily from ${\cal I}_q$ in any way. In fact, as we show shortly, it is not ${\cal L}_q(\omega)$ but energyscales that emerge from it that admit universal scaling near the MBLT. Second, the complex topology of Hilbert-space graphs means it is not amenable towards an explicit identification of the volumescales/lengthscales therein. However, the one-dimensional nature of the (quasi)energy space (see Fig.~\ref{fig:spectral-decomp}) makes ${\cal L}_q(\omega)$, as defined in Eq.~\ref{eq:Lq}, convenient for extracting underlying energyscales.}

With the definition of $\Ll_q(\omega)$ in Eq.~\ref{eq:Lq} and its limiting behaviours at hand, let us discuss the broadbrush features on physical grounds, in both the ergodic and MBL phases.
In the ergodic phase, for $\omega>\omega_\ast$, we expect that the inhomogeneities in $P_n(\omega)$ over $n$ to be washed out such that each $P_n(\omega)\sim n_{\rm seg}^{-1}\sim \omega$ is approximately uniform. We then have $\Ll_q(\omega)\sim \omega^{q-1}$. For $\omega<\omega_\ast$,  $\Ll_q(\omega)$ scales with $\omega$ with a non-universal, anomalous exponent which flows with $\omega$ as different $\omega$ probes the underlying multifractality at different scales. However, as discussed above, as $\omega\to 0$, $\Ll_q(\omega\to 0)\to \mathcal{I}_q \times \omega^0$. This motivates a universal scaling ansatz for $\Ll_q$ in the ergodic phase
\eq{
\Ll_q^{\rm erg}(\omega) = \mathcal{I}_q f_q\left(\frac{\omega}{\omega_\ast}\right);\,\, f_q(x)=\begin{cases} x^{q-1}; & x\gg 1\\ 1; & x\to 0\end{cases}\,.
\label{eq:Lq-erg-scaling}
}
Note that for $\omega>\omega_\ast$, since $\Ll_q(\omega)$ probes the global ergodicity of the eigenstate, there should not be any system-size dependence in this regime. However, in the ergodic phase $\mathcal{I}_q \sim \nh^{-(q-1)}$. It therefore follows that $\omega_\ast \sim \nh^{-1}$ in the ergodic phase with a proportionality factor, $\Lambda_\omega$  defined via $\omega_\ast =\Lambda_\omega \nh^{-1}$. Very deep in an ergodic phase, one expects that the global ergodicity of the wavefunction sets in very quickly due to the high propensity of the wavefunction to have finite and uniform overlap on all $\ket{I}$. On the other hand, as one approaches the MBLT from the ergodic side, the fluctuations in the spectral decomposition grow and the global ergodicity sets in at larger energyscales. As such, one expects the scale $\Lambda_\omega$ to grow as the MBLT is approached and as will be shown later, it diverges at the MBLT.

The situation, however, is very different in the MBL phase. At $\omega$-scales much below the characteristic energyscale, $\Ll_q(\omega)$ picks out only the isolated peaks in the decomposition (see right panel of Fig.~\ref{fig:spectral-decomp}) similar to what happens as $\omega\to 0$.  
Essentially, this means the behaviour of $\Ll_q(\omega)$ at $\omega\to 0$ should persist all the way up to $\omega_\ast$, which automatically implies that there should be an approximate plateau in $\Ll_q(\omega)$ at the value of ${\cal I}_q$ for all $\omega<\omega_\ast$.
At the same time, the normalisation of the wavefunction constrains $\Ll_q(\omega)=1$ for $\omega=2\pi$. 
Consequently, $\Ll_q(\omega)$ has to rise rapidly from its value of ${\cal I}_q\sim \nh^{-\tau_q}$ at $\omega\lessapprox \omega_\ast$ to a value of 1 at $\omega=2\pi$. This suggests that $\Ll_q(\omega)$ does not admit a universal scaling form over the entire range of $\omega$ in the MBL phase as it did in the ergodic phase. Formally, 
\eq{
\Ll_q^{\rm MBL}(\omega) \approx \begin{cases}\mathcal{I}_q; & \omega\ll \omega_\ast \\
\tilde{f}(\omega,L,\Gamma); &\omega \gtrapprox \omega_\ast\end{cases}\,,
\label{eq:Lq-mbl-scaling}
}  
where the function $\tilde{f}$ encodes the non-universal growth. In such a situation, instead of scale-collapsing the entire data, we do so only for the range of the $\omega$ where $\Ll_q(\omega)$ deviates from ${\cal I}_q$ as that is sufficient for us to extract $\omega_\ast$. 
As we will show shortly, $\omega_\ast\sim \nh^{-\mu(\Gamma)}$ with $0<\mu<1$ and $\mu$ increasing with $\Gamma$. 
The latter can be understood as deeper in the MBL phase (smaller $\Gamma$) the peaks in the spectral decomposition get sparser and hence $\omega_\ast$ increases for a given $L$.

As a concrete model for demonstrating the above ideas explicitly, we consider a disordered, Floquet spin-chain~\cite{zhang2016floquet,sierant2023stability} described by the Floquet unitary of the general form mentioned in Eq.~\ref{eq:H-UF-gen} with 
\eq{
\begin{split}
H_{\rm loc} &= T\sum_{i=1}^L[Z_i Z_{i+1} + (h+g\sqrt{1-\Gamma^2}\epsilon_i) Z_i]\,,\\
H_{\rm deloc}&= T g\sum_{i=1}^L X_i\,,
\end{split}
\label{eq:HX-HZ}
}
where $Z_i (X_i)$ denotes the Pauli-$z(x)$ operator on site $i$ and $\epsilon_i\sim \mathcal{N}(0,1)$ are independent Gaussian random numbers. As in Ref.~\cite{zhang2016floquet}, we consider $g=0.9045$, $h=0.809$, and $T=0.8$, for which, within the limits of system sizes accessible there is a putative MBLT at $\Gamma_c\approx 0.3$, with the model in an ergodic(MBL) phase for $\Gamma >(<)\Gamma_c$.

\begin{figure}[t]
\includegraphics[width=\linewidth]{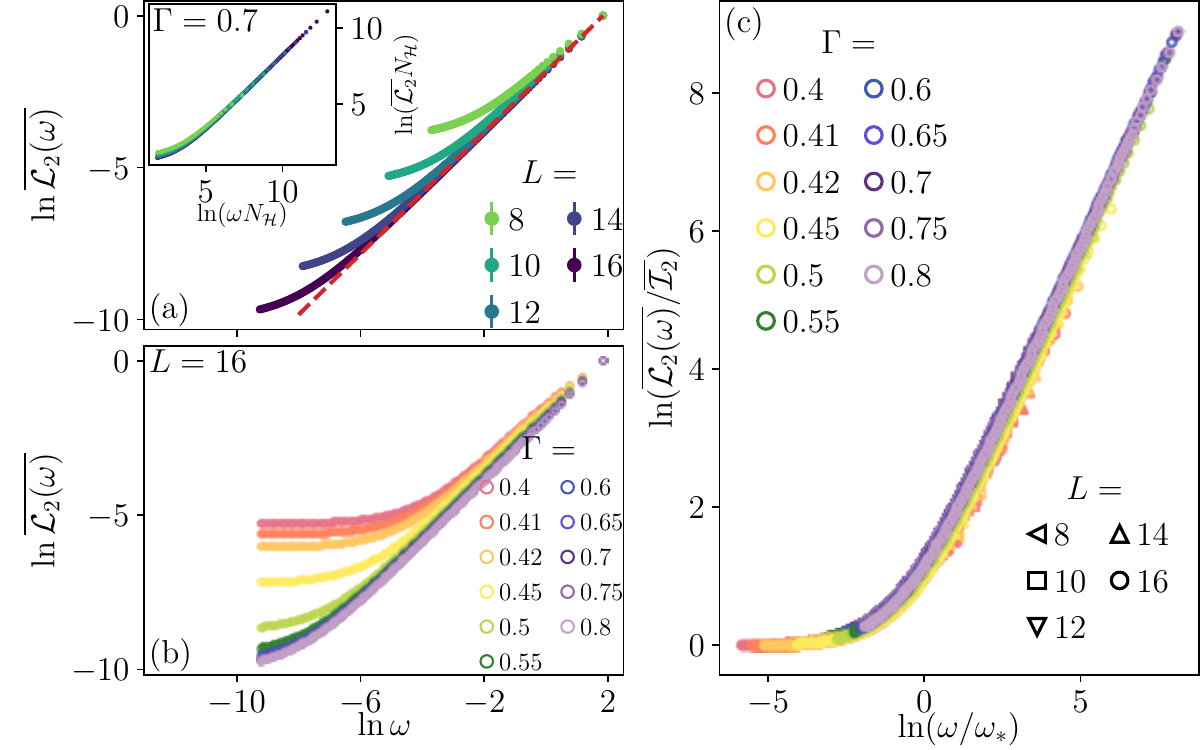}
    \caption{{\bf Results for $\Ll_2(\omega)$ in the ergodic phase.} (a) $\overline{\Ll_2}(\omega)$ as a function of $\omega$ on logarithmic scales for different system sizes, $L$, for $\Gamma=0.7$. The red dashed line denotes the $\omega/2\pi$ line which is the expected asymptotic behaviour. The inset shows the same data but with both $\overline{\Ll_2(\omega)}$ and $\omega$  scaled by $\nh$ showing a perfect collapse for different $L$. (b) $\overline{\Ll_2}(\omega)$ as a function of $\omega$ for different $\Gamma$ in the ergodic phase for $L=16$ shows that $\omega_\ast$ grows as one approaches the MBLT (decreasing $\Gamma$) in the ergodic phase. (c) Evidence for the scaling form in Eq.~\ref{eq:Lq-erg-scaling}. Scaling the $\overline{\Ll_2(\omega)}$ by the IPR, $\overline{\mathcal{I}_2}$, and $\omega$ by $\omega_\ast$ makes the data for all $\Gamma$ (different colours) and all $L$ (different symbols) collapse onto a single curve.}
    \label{fig:ergodic}
\end{figure}

Let us start with the results in the ergodic phase, shown in Fig.~\ref{fig:ergodic}. 
In panel (a) we show $\overline{\Ll_2(\omega)}$ (the overline denotes an average over disorder realisations and eigenstates) as function of $\omega$ for different system sizes $L$ for a representative value of $\Gamma$ in the ergodic phase. The emergence of a characteristic $\omega_\ast$ which depends on $L$ is evident in the data. For $\omega>\omega_\ast$,  $\overline{\Ll_2(\omega)}$ falls exactly on the $\omega/2\pi$ line (red dashed). However as $\omega$ is decreased, the data peels off from this line and the $L$-dependent $\omega$ where this happens can be identified as the $\omega_\ast$. As discussed above, in the ergodic phase, we expect that $\mathcal{I}_q\sim \nh^{-(q-1)}$ and $\omega_\ast\sim \nh^{-1}$. That this is indeed the case is confirmed by the perfect collapse of the data when both $\overline{\Ll_2(\omega)}$ and $\omega$ are scaled by $\nh$ as shown in the inset. The data in panel (b) shows that $\omega_\ast$ for a given $L$ indeed increases as $\Gamma$ is decreased towards the MBLT. In panel (c), we plot the data for a range of $\Gamma$ and $L$ by scaling $\overline{{\cal L}_2 (\omega)}$ with the average IPR, $\overline{{\cal I}_2}$, and $\omega$ by $\omega_\ast$ which depends on $\Gamma$ and $L$; this leads to a remarkably good collapse of the data for all $L$ and $\Gamma$ onto a single curve hence providing evidence for the validity of the scaling form in Eq.~\ref{eq:Lq-erg-scaling} throughout the ergodic phase. The scaling collapse for each $\Gamma$ separately is shown in the Supp. Matt.~\cite{supp}.

\begin{figure}[t]
    \includegraphics[width=\linewidth]{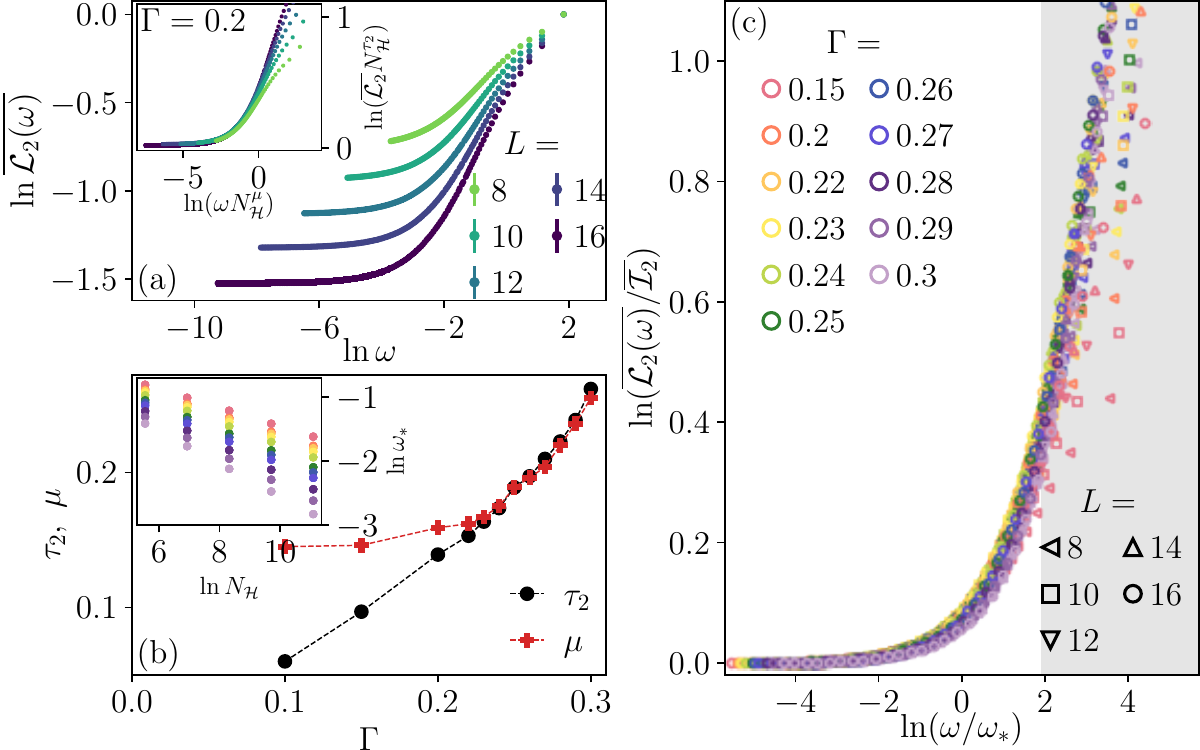}
    \caption{{\bf Results for $\Ll_2(\omega)$ in the MBL phase.} (a) $\overline{\Ll_2}(\omega)$ as a function of $\omega$ for different system sizes, $L$, for $\Gamma=0.2$. The inset shows the same data but with $\overline{\Ll_2(\omega)}$ scaled by $\nh^{-\tau_2}$ and $\omega$  scaled by $\nh^{-\mu}$ showing a collapse for different $L$.  (b) The exponent $\mu$ as a function of $\Gamma$ and its comparison to the fractal exponent $\tau_2$ from the IPRs. The inset shows $\ln\omega_\ast$ as a function of $\ln\nh$ for different $\Gamma$ (see legend in panel (c)). (c) Collapse the data for $\overline{\Ll_2(\omega)}$ scaled by the IPR versus $\omega$ scaled by $\omega_\ast$ for several values of $L$ and $\Gamma$ in the MBL phase. The collapse (also in the inset to (a)) is seen only for the flat regime and the regime where $\overline{\Ll_2(\omega)}/\overline{{\cal I}_2(\omega)}$ begins to deviate from unity. The grey shaded region is a guide to the eye for the region where we do not expect the scaling to be valid.}
    \label{fig:mbl}
\end{figure}

We will return to the results for the scaling of $\omega_\ast$ as $\Gamma\to\Gamma_c^+$ but before that, we discuss $\Ll_2(\omega)$ in the MBL phase. 
The results are shown in Fig.~\ref{fig:mbl}. The data in panel (a) shows that $\Ll_2(\omega)$ sticks approximately to its $\omega\to 0$ value, ${\cal I}_2$, for an extended range of $\omega$ above which it takes off towards unity, its value at $\omega=2\pi$. 
The inset shows the same data but ${\Ll}_2(\omega)$ scaled by $\nh^{-\tau_2}$ and $\omega$ scaled by $\nh^{-\mu}$ with $0<\mu<1$.
While such a scaling does not collapse the entire curves, it does so for the the regime of $\omega$ where $\Ll_2(\omega)$ begins to deviate from the flat region and hence, identifies $\omega_\ast$. 
In panel (c), we show data for several $L$ and $\Gamma$ with $\Ll_2(\omega)$ scaled with the IPR and $\omega$ scaled with $\omega_\ast$ (the data for each $\Gamma$ separately is shown in the Supp. Matt.~\cite{supp}). As in the inset to panel (a), the data shows an excellent collapse in the region of interest. The exponent $\mu$ as a function of $\Gamma$ is shown in panel (b), which decreases on decreasing $\Gamma$ indicating that $\omega_\ast\sim \nh^{-\mu}$ does indeed grow as one goes deeper into the MBL phase. 
The inset shows $\ln\omega_\ast$ against $\ln\nh$, straight line fits to which were used to obtain $\mu$.
We also find that this behaviour of $\omega_\ast$ continues all the way upto and at the critical point with $\omega_{\ast,c}\equiv\omega_\ast(\Gamma_c)\sim \nh^{-\mu_c}$ with our best estimate of the critical point being $\Gamma_c\approx 0.3$.
This is again indicative of the fact that the MBLT point is a part of the MBL phase itself.
It is also worth noting in panel (b) that the fractal exponent $\tau_2$ as function of $\Gamma$ is remarkably close to $\mu$ except very deep in the MBL phase; we shall return to a physical understanding of this later.

\begin{figure}[!b]
\includegraphics[width=0.48\linewidth]{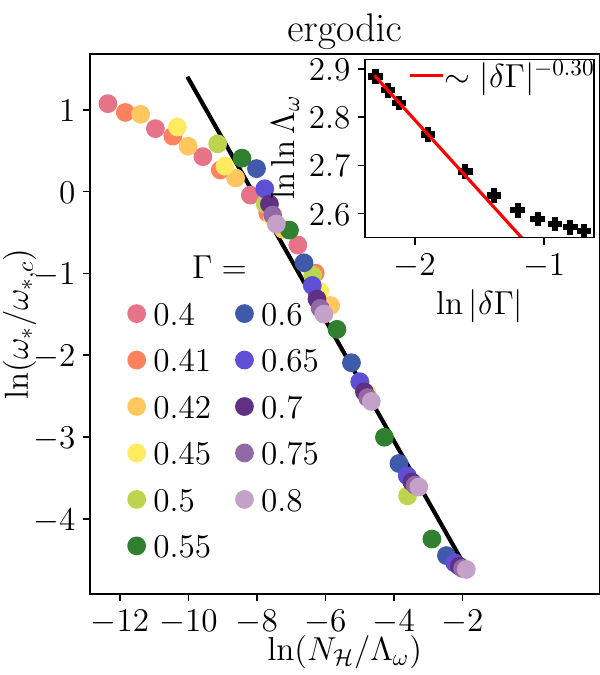}
\includegraphics[width=0.48\linewidth]{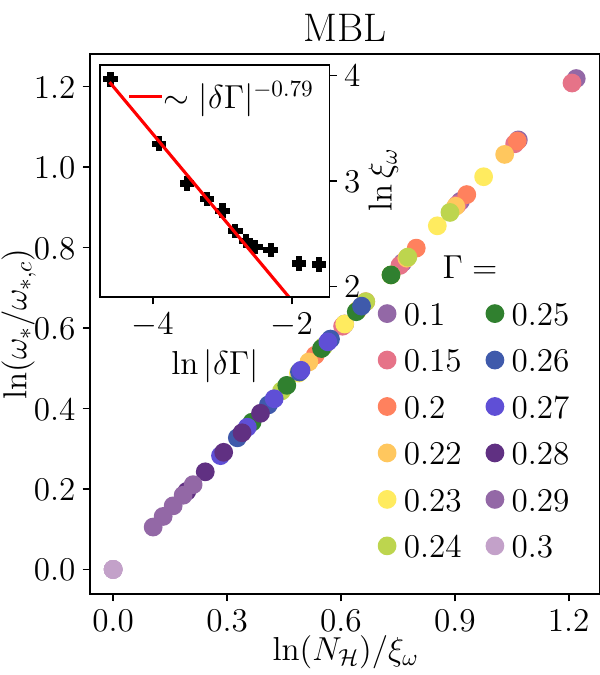}
    \caption{{\bf Scaling of $\omega_\ast$.} (a) Data for $\omega_\ast/\omega_{\ast,c}$ for several $L$ and $\Gamma$ plotted as a function of $\nh/\Lambda_\omega$ in the ergodic phase (left) and as a function of $\ln\nh/\xi_\omega$ in the MBL phase (right) collapses onto common curves evincing the volumic and linear scaling in Eq.~\ref{eq:omegastar-scaling}. The inset in (a) shows the divergence of the correlation energy-volume $\Lambda_\omega$ with $\delta\Gamma$ as the MBLT is approached with an essential singularity whereas the inset in (b) shows the power-law divergence of the correlation energy length $\xi_\omega$ as the MBLT is approached from the MBL side.}
    \label{fig:omegastar}
\end{figure}

The numerical results discussed so far provide compelling evidence for the first central result of the work; namely the identification of an emergent characteristic energyscale, $\omega_\ast$, which scales $\sim\nh^{-1}$ in the ergodic phase and $\sim \nh^{-\mu}$ with $0<\mu(\Gamma)<1$ in the MBL phase. 
We next turn towards numerical results for the second key point of the work, namely, the critical behaviour of $\omega_\ast$.
It is remarkable to note that the scaling of $\omega_\ast$ with $\nh$ bears stark similarities to that of the IPRs~\cite{mace2019multifractal,roy2021fockspace}. 
It is therefore natural to expect a scaling ansatz for $\omega_\ast$ near the MBLT similar to those for IPRs~\cite{mace2019multifractal,roy2021fockspace,supp},
\eq{
\omega_\ast = \omega_{\ast,c}\times
\begin{cases}
{\cal F}_\omega^{\rm vol}\left(\frac{\nh}{\Lambda_\omega}\right);& \Gamma>\Gamma_c~{\rm (ergodic)}\\
{\cal F}_\omega^{\rm lin}\left(\frac{\ln \nh}{\xi_\omega}\right);& \Gamma<\Gamma_c~{\rm (MBL)}
\end{cases}\,.
\label{eq:omegastar-scaling}
}
The scaling ans\"atz in Eq.~\ref{eq:omegastar-scaling}, which were originally motivated by the critical behaviour of IPRs across Anderson transitions in high-dimensional graphs~\cite{garciamata2017scaling,garciamata2022critical} imply that the correction to the critical behaviour of $\omega_\ast$ as one moves away from the critical point towards the ergodic phase satisfies a {\it volumic} scaling form with a correlation volume-scale in energy, $\Lambda_\omega$, whereas on moving towards the MBL side, it satisfies a {\it linear} scaling form with a characteristic correlation lengthscale in energy, $\xi_\omega$. The results in Fig.~\ref{fig:omegastar} show that the volumic and linear scaling ans\"atz in Eq.~\ref{eq:omegastar-scaling} indeed provide an excellent description of the data. In the ergodic phase, for $\nh\gg \Lambda_\omega$, the fully ergodic $\nh^{-1}$ scaling of $\omega_\ast$ sets in; the resulting asymptotic behaviour of $\omega_\ast/\omega_{\ast,c}\sim \nh^{\mu_c-1}$ is found to be in good agreement with the numerical data as shown by the black line in Fig.~\ref{fig:omegastar} (left). 
The insets to the two panels in Fig.~\ref{fig:omegastar} show the behaviour of $\Lambda_\omega$ in the ergodic phase and $\xi_\omega$ in the MBL phase with $\delta\Gamma \equiv \Gamma-\Gamma_c$. 
The results show that in the ergodic phase $\Lambda_\omega$ diverges with an essential singularity on approaching the MBLT,
\eq{
\Lambda_\omega\sim \exp\left[\frac{c}{|\delta \Gamma|^\beta}\right]; \quad \beta\approx 0.3\,,
\label{eq:Lambda-om}
}
whereas in the MBL phase, the $\xi_\omega$ diverges as a power-law on approaching the MBLT,
\eq{
\xi_\omega \sim |\delta\Gamma|^{-\nu};\quad \nu\approx 0.79\,,
\label{eq:xi-om}
} 
thus suggesting the Kosterlitz-Thouless-like scenario for the MBLT.
One remarkable aspect of the results in Eq.~\ref{eq:Lambda-om} and Eq.~\ref{eq:xi-om} is that they show that the scaling theory of the MBLT based on $\omega_\ast$ as discussed above bears stark resemblances with a scaling theory based on the eigenstate IPRs not only at a qualitative level but also quantitatively.
The IPRs also admit a volumic and linear scaling in the ergodic and MBL phases respectively~\cite{supp} with the correlation volume, $\Lambda$, in the ergodic phase diverging as $\Lambda\sim \exp[c'/|\delta \Gamma|^{0.37}]$ and the correlation length in the MBL phase, $\xi$, diverging as $\xi \sim |\delta \Gamma|^{-0.74}$~\cite{supp}.

The quantitative similarity between the scaling theories based on $\omega_\ast$ and that based on $\overline{{\cal I}_2}$, while {\it a priori} unexpected and not obvious, leads to useful insights about the connections between the two theories. 
In the ergodic phase, the spectral decomposition can be viewed as a set of fractals of extent $\omega_\ast=\Lambda_\omega\nh^{-1}$ stacked next to each other such that on zooming out to scales $\omega>\omega_\ast$, the decomposition appears ergodic. This is exactly analogous to the idea that the eigenstates on the Fock-space graph can be interpreted as a repeated {\it tiling} on the Fock-space graph of multifractal states of size $\Lambda$~\cite{garciamata2017scaling,mace2019multifractal,roy2021fockspace,garciamata2022critical}. 
It therefore leads to the understanding of why $\Lambda_\omega$ and $\Lambda$ exhibit quantitatively similar scaling at the MBLT.

In the MBL phase, the quantitative similarity between $\mu$ and $\tau$ can be justified via the  extremely naive picture that the spectral decomposition has $O(\nh^{\tau_2})$ peaks and they are separated, on average, by (quasi)energy $\sim \nh^{-\tau_2}$. This automatically implies that $\Ll_2(\omega)$ probes the individual peaks as long as $\omega \lesssim \nh^{-\tau_2}$ and deviates from this behaviour beyond that; hence $\omega_\ast\sim \nh^{-\mu}$ with $\mu\approx \tau_2$. 
It is worth mentioning at this stage an alternative viewpoint for understanding the existence of $\omega_\ast$ and its scaling with $\nh$. It was shown for multifractal states in general~\cite{altshuler2016multifractal} that on computing the local density of states (LDoS) with a line-broadening $\eta$, the scaling of the typical LDoS with $\eta$ saturated for $\eta<\eta_\ast\sim\nh^{-z} \gg\nh$ ($0<z<1$) and the typical LDoS depended solely on the Hilbert-space dimension . In this case, $\eta$ can be interpreted as setting the scale at which the LDoS is probed analogous to $\omega$. This immediately implies the presence of a $\omega_\ast$ such that $\Ll_q(\omega)$ depends solely on $\nh$ for $\omega<\omega_\ast$ and on both $\nh$ and $\omega$ otherwise.

Heuristically, the scale $\omega_\ast$ can also be related to rare resonances in the MBL phase. 
{Here, the term resonance refers to the phenomenology that an eigenstate of $H$ or $U_F$ may develop significant support on several eigenstates of $H_{\rm loc}$; and these eigenstates of $H_{\rm loc}$ may be substantially different from the eigenstate of $H_{\rm loc}$ to which the said eigenstate of the full $H$ is smoothly connected to (via a finite-depth local unitary) in the limit of infinitely strong disorder.}
The probablity that a given state participates in a resonace involving $r$ spins is typically exponentially small in $r$, $p_{\rm res}(r) \sim e^{-r/\zeta}$. At the same time, the number of such possible resonances is $N_r=\binom{L}{r}$. The total number of resonances that the a state participates in typically is thus $N_{\rm res}=\sum_{r=0}^Lp_{\rm res}N_r\sim \nh^{\gamma}$ with $\gamma = \ln(1+e^{-1/\zeta})/\ln2$. The typical separation in $\omega$ between these resonances is $2\pi/N_{\rm res}\sim \nh^{-\gamma}$ suggesting that $\omega_\ast\sim \nh^{-\gamma}$ and therefore $\mu=\gamma$. Note that this is consistent with the behaviour on approaching the MBLT from the MBL side, the scale $\zeta$ remains finite at the MBLT whereas upon crossing into the ergodic phase, it diverges indicating a proliferation of the resonances.

To summarise, by probing the multifractal properties of the spectral decomposition of eigenstates at different energyscales, we identified characteristic energyscales in either phase which show universal behaviour near the MBLT. The critical scaling of these scales strongly resembles that of eigenstate IPRs and thus admits a scenario consistent with a Kosterlitz-Thouless-type scaling at the MBLT. Before we close, a few comments regarding some of the questions and directions for future research raised by this work are in order.

The results presented here, for locally interacting systems, should be contrasted with the ideas  presented in Ref.~\cite{monteiro2021quantum} for non-locally interacting systems where it was suggested that eigenstates are spread over a Fock-space energy shell, $\kappa$, in an ergodic fashion although the shell itself fills up only a fraction of the Fock space. That would imply that $\Ll_q(\omega)\sim\omega^{q-1}$ for $\omega\lesssim\kappa$ and $\approx 1$ otherwise. We however see no such behaviour in our case; a systematic study of $\Ll_q(\omega)$ for such non-locally interacting models is therefore important for clarifying this interesting distinction.

One useful fallout of the analysis presented in this work is that the space of quasienergies $\theta_I$ serves as a useful one for studying the multifractal properties of the eigenstates at different scales. Given that the MBL eigenstates are genuinely multifractal, analogous to single particle wavefunctions at Anderson transitions~\cite{mirlin2000multifractality,evers2008anderson}, the dynamical eigenstate correlations similar to those considered in Ref.~\cite{chalker1988scaling,chalker1990scaling} but with real space replaced by $\theta$-space is of immediate interest. Whether the resulting scalings are consistent with conformal invariance and homogeneity under combined transformations of eigenenergies and $\theta$ will lead to fundamental insight about the nature of MBL eigenstates.

\begin{acknowledgements}
We thank A. Kundu and A. Lazarides for useful comments on an earlier version of the manuscript.
We acknowledge support from SERB-DST, Government of India under Grant No. SRG/2023/000858, from the Department of Atomic Energy, Government of India, under Project No. RTI4001, and from an ICTS-Simons Early Career Faculty Fellowship via a grant from the Simons Foundation (Grant No. 677895, R.G.)
\end{acknowledgements}

\bibliography{refs}

\clearpage

\setcounter{equation}{0}
\setcounter{figure}{0}
\setcounter{page}{1}
\renewcommand{\theequation}{S\arabic{equation}}
\renewcommand{\thefigure}{S\arabic{figure}}
\renewcommand{\thesection}{S\arabic{section}}
\renewcommand{\thepage}{S\arabic{page}}

\onecolumngrid

\begin{center}
{\bf {{Supplementary Material: Spectral Multifractality and Emergent Energyscales at the Many-Body Localisation Transition}}}\\
\medskip

Sthitadhi Roy\\
{\it {\small International Centre for Theoretical Sciences, Tata Institute of Fundamental Research, Bengaluru 560089, India}}
\end{center}

This supplementary material contains additional numerical data for the quantities discussed in the main text

\section{I. Scaling of Inverse Participation Ratios}

\begin{figure}[!b]
\subfloat[Ergodic phase \label{fig:iprdeloc}]{%
  \includegraphics[width=0.45\linewidth]{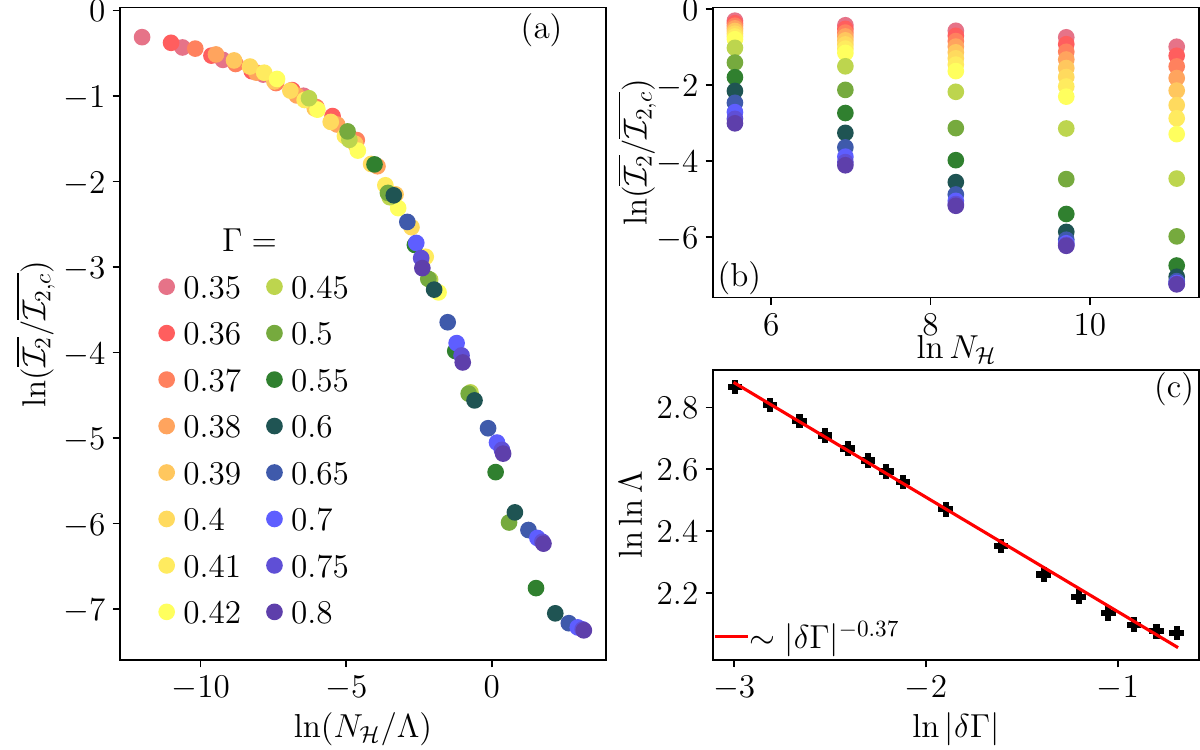}%
}\hfill
\subfloat[MBL phase\label{fig:iprloc}]{%
  \includegraphics[width=0.455\linewidth]{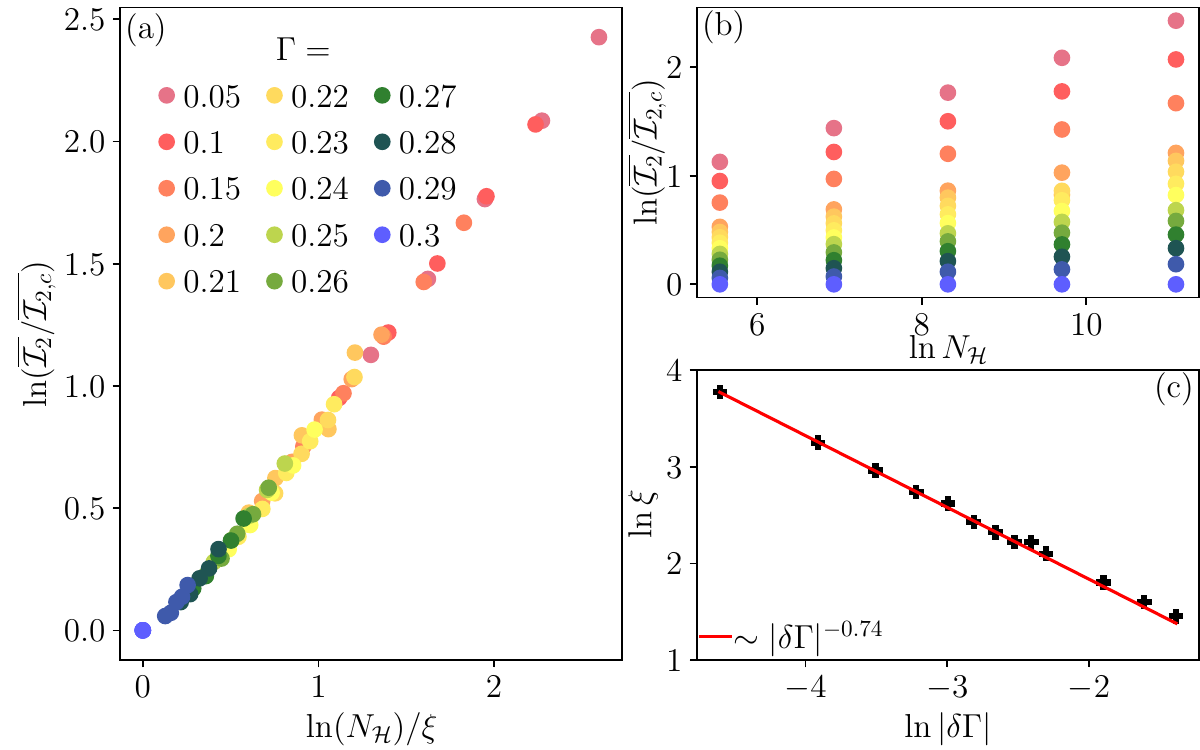}%
}
\caption{{\bf Scaling of the IPRs in the ergodic phase (A) and in the MBL phase (B).} Panels (a) in both sets of the data show the data for $\overline{{\cal I}_2}/{\overline{{\cal I}_{2,c}}}$ as a function of the respective scaling variable, $\nh/\Lambda$ in the ergodic phase and $(\ln \nh)/\xi$ in the MBL phase. The data as a function of $\nh$ (without any scaling) is shown in panels (b). The data in panels (c) show the divergence of $\Lambda$ with an essential signularity in the ergodic phase and that of $\xi$ as a power-law in the MBL phase. }
\label{fig:iprscaling}
\end{figure}

One of the central results discussed in the main text was the quantitative similarities of the scaling of $\omega_\ast$ near the MBLT with that of the IPRs. To provide numerical evidence for this, in this section, we shows the data for the scaling of IPRs. Taking cue from Refs.~\cite{mace2019multifractal,roy2021fockspace}, we use the scaling ansatz
\eq{
\overline{{\cal I}_2} = \overline{{\cal I}_{2,c}}\times
\begin{cases}
{\cal G}^{\rm vol}\left(\frac{\nh}{\Lambda}\right);& \Gamma>\Gamma_c~{\rm (ergodic)}\\
{\cal G}^{\rm lin}\left(\frac{\ln \nh}{\xi}\right);& \Gamma<\Gamma_c~{\rm (MBL)}
\end{cases}\,,
\label{eq:ipr-scaling}
}
where $\Lambda$ and $\xi$ denote Fock-space correlation volumes and correlation lengths respectively on the ergodic and MBL side of the MBLT. The results in panels (a) in Fig.~\ref{fig:iprdeloc} for the ergodic phase and Fig.~\ref{fig:iprloc} in the MBL phase  show that the data does indeed satisfy the scaling ansatz in Eq.~\ref{eq:ipr-scaling} remarkably well. 
The results in panel (c) of Fig.~\ref{fig:iprdeloc} shows that the correlation volume $\Lambda$ diverges as essential singularity with 
\eq{
\Lambda \sim \exp[-c/|\delta \Gamma|^{\beta}]\,;\quad \beta\approx0.37\,.
}
The exponent $\beta$ is quite close to that obtained from the divergence of the correlation volumescale in energy, $\Lambda_\omega$; for the latter we had obtained $\beta\approx 0.3$ (see Fig.~\ref{fig:omegastar} (left)).
The analogous figure for the MBL phase, panel (c) of Fig.~\ref{fig:iprloc} shows that the lengthscale $\xi$ diverges with a power-law as 
\eq{
\xi\sim |\delta\Gamma|^{-\nu}\,;\quad \nu\approx 0.74\,,
}
where again the exponent $\nu$ is remarkably close to that same exponent obtained for the divergence of $\xi_\omega$ for which we had $\nu\approx 0.79$ (see Fig.~\ref{fig:omegastar} (right)).

\begin{figure*}[!]
\includegraphics[width=0.98\linewidth]{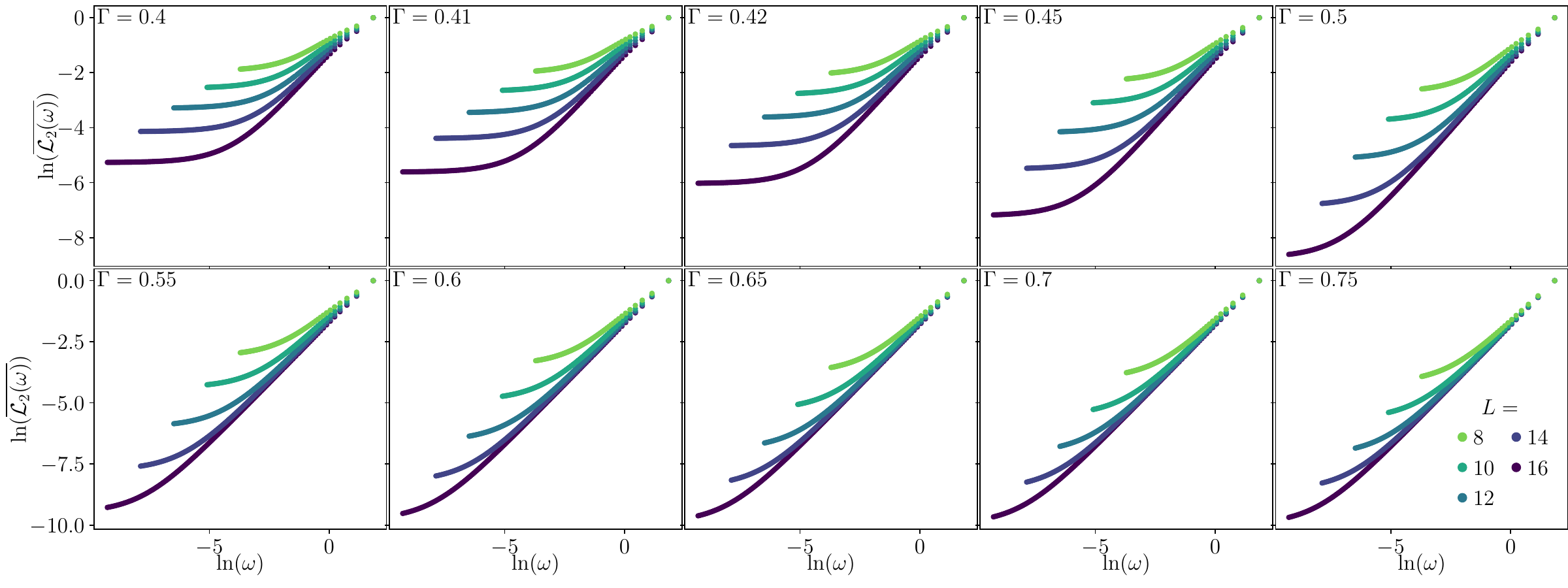}
\caption{The raw data for $\overline{\Ll_2(\omega)}$ versus $\omega$ for several values of $\Gamma$ in the ergodic phase of the disordered, Floquet spin-1/2 chain \eqref{eq:HX-HZ}. Different colours correspond to different $L$ as mentioned in the legend whereas each panel corresponds to a $\Gamma$ value as mentioned in the top left corner of the panel.}
\label{fig:erg-all}
\end{figure*}

\begin{figure*}[!]
\includegraphics[width=0.98\linewidth]{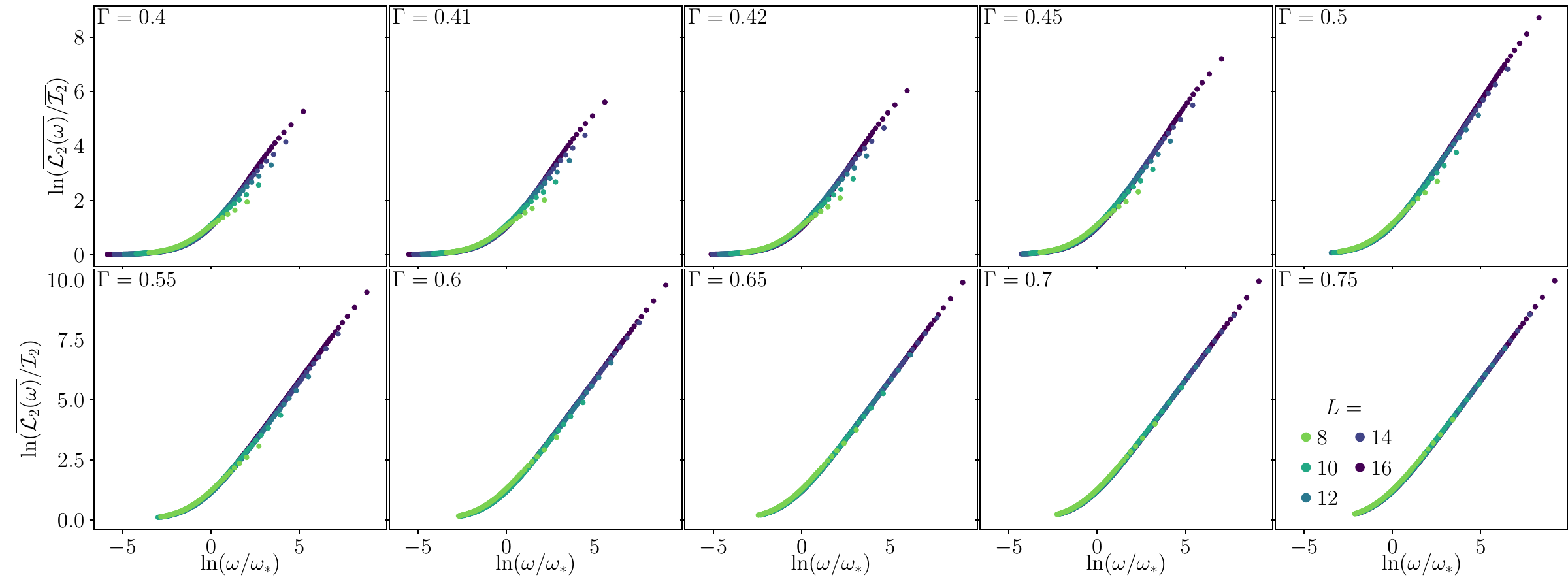}
\caption{The same set of data as in Fig.~\ref{fig:erg-all} but with the $\overline{\Ll_2(\omega)}$ scaled by the IPR, $\overline{{\cal I}_2}$, and $\omega$ scaled by the characteristic energyscale $\omega_\ast$.}
\label{fig:erg-all-scaled}
\end{figure*}

\section{II. Additional data for $\Ll_2(\omega)$}
\subsection{II. A. Ergodic phase}
In Fig.~\ref{fig:erg-all}, we show the raw data for $\overline{\Ll_2(\omega)}$ versus $\omega$ for several values of $\Gamma$ in the ergodic phase. The figure is analogous to Fig.~\ref{fig:ergodic}(a) with each panel corresponding to a particular value of $\Gamma$. Note that the value of $\omega$ at which the data deviates from the $\sim\omega$ line, which we identify as $\omega_\ast$ clearly goes down for a fixed $L$ as we move deeper into the ergodic phase. The scale-collapsed data, where $\overline{\Ll_2(\omega)}$ is scaled by the IPR, $\overline{{\cal I}_2}$, and $\omega$ is scaled by the characteristic energyscale $\omega_\ast$ is shown in Fig.~\ref{fig:erg-all-scaled}. Note that Fig.~\ref{fig:ergodic}(c) is essentially obtained by overlaying all the panels in  Fig.~\ref{fig:erg-all-scaled}; the sole purpose of the latter is show the remarakably good quality of scaling collapse for each value of $\Gamma$.

\subsection{II. B. MBL phase}
\begin{figure*}[!]
\includegraphics[width=0.98\linewidth]{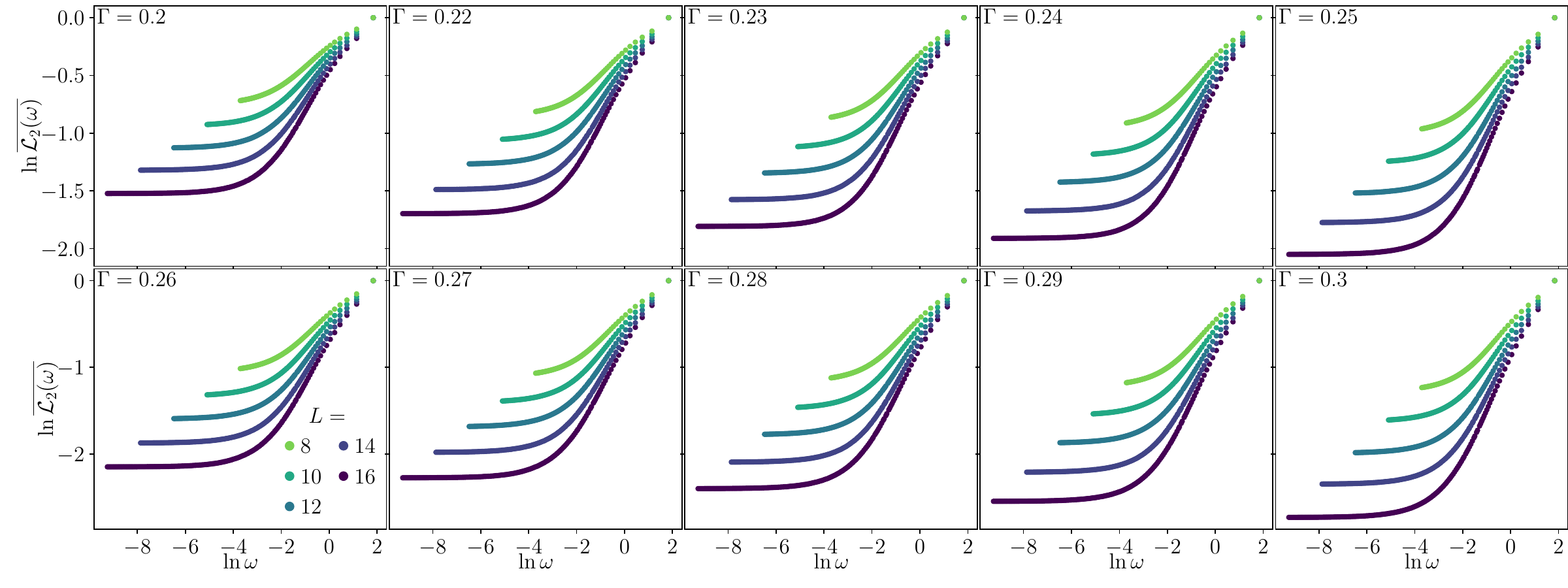}
\caption{The raw data for $\overline{\Ll_2(\omega)}$ versus $\omega$ for several values of $\Gamma$ in the MBL phase of the disordered, Floquet spin-1/2 chain \eqref{eq:HX-HZ}. Different colours correspond to different $L$ as mentioned in the legend whereas each panel corresponds to a $\Gamma$ value as mentioned in the top left corner of the panel.}
\label{fig:mbl-all}
\end{figure*}

\begin{figure*}[!]
\includegraphics[width=0.98\linewidth]{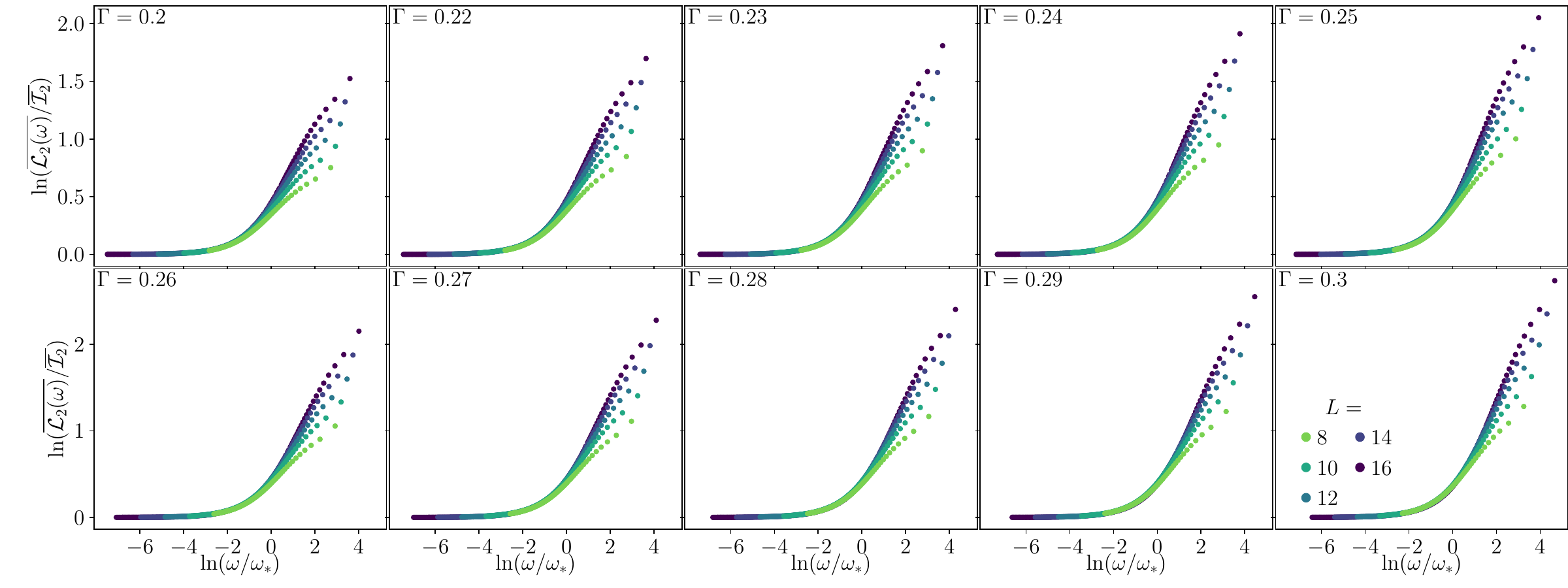}
\caption{The same set of data as in Fig.~\ref{fig:mbl-all} but with the $\overline{\Ll_2(\omega)}$ scaled by the IPR, $\overline{{\cal I}_2}$, and $\omega$ scaled by the characteristic energyscale $\omega_\ast$. As discussed in the main text, the entire curve does not collapse. However, the part of the graph where $\overline{\Ll_2(\omega)}$ begins to deviate from $\overline{{\cal I}_2}$ collapses for different $L$.}
\label{fig:mbl-all-scaled}
\end{figure*}

Plots analogous to Fig.~\ref{fig:erg-all} and Fig.~\ref{fig:erg-all-scaled}, but for the MBL phase are shown in Fig.~\ref{fig:mbl-all} and Fig.~\ref{fig:mbl-all-scaled}. Throughout the MBL phase, one can see in Fig.~\ref{fig:mbl-all} the flat behaviour of $\Ll_2(\omega)$ at ${\cal I}_2$ for $\omega<\omega_\ast$ beyond it which it deviates. Scaling $\omega$ with $\omega_\ast$, as in Fig.~\ref{fig:mbl-all-scaled} collapses this flat region and the deviation from it, which is sufficient for the identification of $\omega_\ast$, but does not collapse the entire curve as alluded to in the main text.
 
\end{document}